# Magnesium Diboride Flexible Flat Cables for Cryogenic Electronics

Chris S. Yung and Brian H. Moeckly

*Abstract*—Magnesium diboride ($MgB_2$) thin films are a potential alternative to low-temperature superconductors (LTS) due to a higher critical temperature ($T_c$) of approximately 39 K. The reactive evaporation deposition technique also affords relatively simple growth of $MgB_2$ films on flexible substrates compared to high-temperature superconductors (HTS). We have designed and fabricated a cable architecture consisting of $MgB_2$ traces on flexible yttria-stabilized zirconia (YSZ) compatible with commercially available connectors or direct wirebonds. Key performance metrics such as critical current density ($J_c$) and $T_c$ are measured and compared. We discuss thermal conductivity and passivation schemes for these cables.

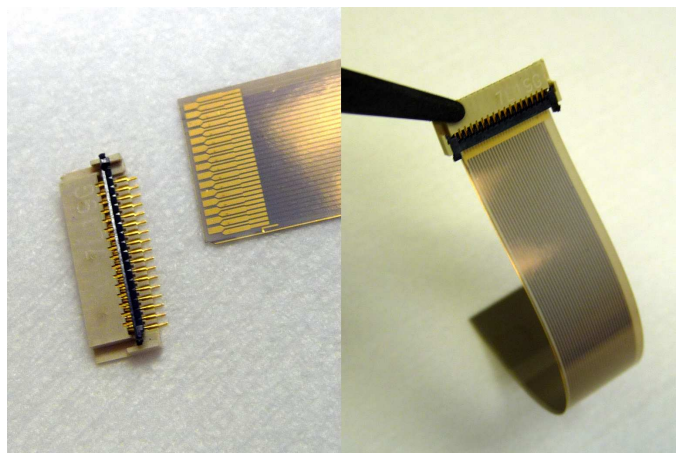

Fig. 1. A flexible flat cable with 33 traces of 200-μm-wide, 300-nm-thick $MgB_2$ deposited on a flexible YSZ tape. The cable is mated to a Hirose FFC/FPC connector. The substrate size is 1 cm × 10 cm × 50 μm.

## I. INTRODUCTION

SINCE the discovery of superconductivity in $MgB_2$ [1], thin film deposition of this material has become realizable on a variety of substrates [2]. The most successful growth techniques include hybrid physical-chemical vapor deposition [3] and reactive evaporation [4]. The advent of these deposition methods along with a relatively high $T_c$ of 38 to 39 K has made $MgB_2$ a considered alternative to conventional low-temperature superconductors such as Nb, Al, and NbN. The reactive evaporation technique also enables growth of $MgB_2$ on non-epitaxial, large-area substrates including flexible dielectrics. In addition, buffer layers are usually not required, and the growth temperature is lower than that required for HTS materials.

Potential thin film electronics technologies utilizing $MgB_2$ include Josephson junctions [5], superconducting single photon detectors [6], bolometers [7], kinetic inductance detectors [8], and flexible cryogenic interconnects for LTS computing [9], [10]. $MgB_2$ is especially appealing since its low resistivity and high $T_c$ values have been shown to be preserved when thin films are deposited on non-single-crystalline substrates, thereby obviating the need for substrate preparation or epitaxial growth. For example, $MgB_2$ thin films have been grown on substrates ranging from polyimide [11], [12] to stainless steel [4], expanding the field of flexible superconducting electronics to more commercially viable substrates. Here we discuss the application of $MgB_2$ thin films for ribbon cables or flexible flat cables/flexible printed circuits (FFC/FPC) for low-temperature technologies.

There exists a need for low-thermal-conductivity high-density interconnects for a multitude of low-temperature applications. Existing dc electrical connections are often routed with circular wire looms of low thermal conductivity alloys such as Manganin, phosphor-bronze, copper-nickel, or even stainless steel. Due to the Wiedemann-Franz law, the aforementioned alloys limit the thermal current by possessing a proportionally high electrical resistance. These wire looms, while advantageous in terms of ease of use and low thermal conductivity, exhibit high electrical resistance, are not wirebond compatible (i.e., solder connections are required), and must be cryogenically thermalized by potting them in epoxy or GE varnish. These disadvantages can be overcome by use of a flat geometry. By utilizing a planar configuration, individual traces can be wirebonded, thereby allowing submicron size devices to be interfaced via FFC to sensitive readout electronics which may also be submicron in scale. FFCs exhibit lower heat dissipation than circular wire due to a superior surface-area to volume ratio. Heat sinking of the flat cable is a simple matter of clamping it between two flat pieces of metal. An example of the use of FFCs in low-temperature applications is Cu-Ni flex cable in the imaging array wiring for transition-edge sensors (TES) operating at 100 mK [13]. The use of superconducting traces instead of a resistive alloy will

Manuscript received August 3, 2010. This work was supported by the Office of Naval Research contract N00014-07-1-0352.
The authors are with Superconductor Technologies, Inc., Santa Barbara, CA 93111 USA (e-mail: bmoeckly@suptech.com).

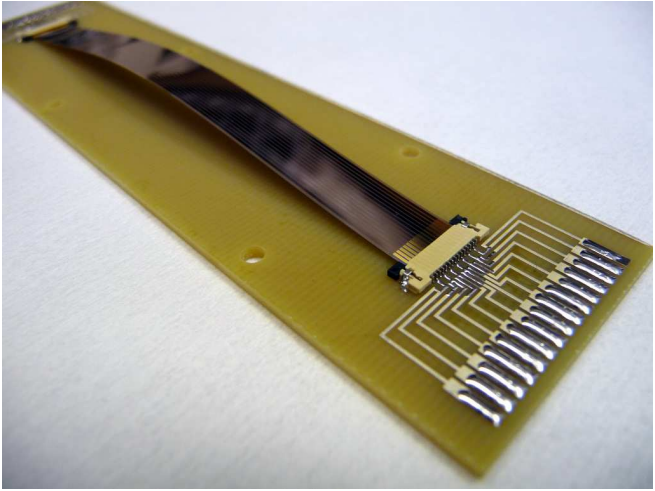

Fig. 2. Circuit board setup with surface mounted JST connectors for cryogenic measurement of the FFCs. The MgB$_2$/YSZ tape is 0.8 cm wide × 10 cm long. The JST connectors support 15 MgB$_2$ traces which are 0.35 mm wide on a 0.5 mm pitch. The connectors are wirebonded to the circuit board.

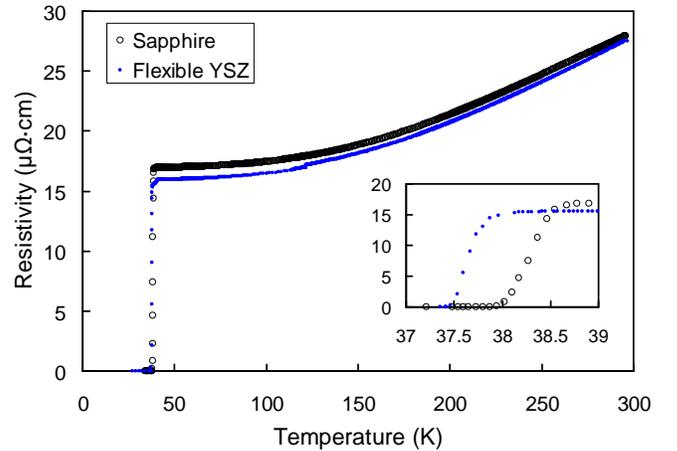

Fig. 3. Resistivity as a function of temperature for MgB$_2$ deposited simultaneously on 1 cm × 1 cm substrates of $r$-plane Al$_2$O$_3$ and flexible YSZ. The films are 100 nm thick and were deposited at 550º C. $T_c$ for the film on $r$-plane sapphire is 37.9 K and the value on flexible YSZ is 37.4 K.

further reduce the overall thermal conductivity of the cable.

In this paper we present measurements of a ribbon cable consisting of superconducting MgB$_2$ traces on flexible yttria-stabilized zirconia. These cables are capable of operating at 30 to 35 K or lower with commercially available FFC/FPC connectors and can support a critical current greater than 100 mA per trace at 4.2 K.

## II. DESIGN AND FABRICATION

### A. Reactive Evaporation of MgB$_2$

We deposit MgB$_2$ thin films via a reactive evaporation technique incorporating a custom heater geometry [4]. Substrates rotate on a circular platen inside a quasi-blackbody radiative heater and are alternatively exposed to B flux deposited in vacuum and Mg vapor localized within the heater. Our present platen is capable of holding three 50-cm-diameter wafers or multiple 1 cm × 10 cm tapes. For this work we deposited MgB$_2$ on flexible YSZ tapes at a temperature of 550 ºC and a growth rate of ~1.2 Å/s. The substrates were 50 μm thick and the films were 100 nm to 300 nm thick. Single-crystal 1 cm × 1 cm $r$-plane Al$_2$O$_3$ witness samples were included with the YSZ tapes for measurement comparison.

### B. Fabrication of Flexible Flat Cables

Flexible YSZ (commercially available as Ceraflex) was chosen as the FFC substrate due to a high working temperature (>2000 ºC) and a minimum bend radius of 8 mm. Using standard photolithography, a photoresist etch mask was used to define MgB$_2$ traces parallel to the tape length with widths of 200 or 350 μm. Samples were patterned by Ar ion milling through the MgB$_2$ and over-etching into the YSZ substrate by 200 nm to ensure electrical isolation between traces. Metallization of the end connections was performed via a photolithographic lift-off procedure following ion sputtering of Ti/Ag/Au. A layer of Kapton tape was applied to the back of the YSZ to increase the overall thickness of the ribbon cable to 0.1 mm in order to meet the thickness requirement of the FFC connectors. The resistance between MgB$_2$ traces was measured with an ohmmeter at room temperature and exceeded 15 MΩ for all traces. A completed MgB$_2$ FFC under flexure is shown in Fig. 1. The 33 MgB$_2$ traces are 200 μm wide and 300 nm thick. The witness samples of MgB$_2$ grown on $r$-plane Al$_2$O$_3$ were patterned and ion milled in a similar fashion to allow 4-point measurements of MgB$_2$ microbridges of widths 1, 2, and 5 μm.

### C. Connectorization

Commercially available FFC/FPC connectors from JST and Hirose were chosen due to their compact size and ease of use. The zero-insertion-force connectors lock the inserted FFC into place by means of a movable plastic clamp. The JST connectors (part no. 15FHZ-SM1-S-TB) support a maximum current of 0.5 A with a contact resistance of 40 mΩ, for 15 circuits of 0.35 mm width on a 0.5 mm pitch. The Hirose connectors (part no. FH23-33S-0.3SHW(05)) support a maximum current of 0.3 A with a contact resistance of 100 mΩ, for 33 circuits of 0.2 mm width (0.1 mm wide in connector region) on a 0.6 mm pitch. MgB$_2$ traces on flexible YSZ tapes were fabricated for both types of connectors; however, for this study only the JST style connectors were measured. Wirebonding to the metallization pads proved to be straightforward.

## III. ELECTRICAL MEASUREMENTS

### A. Cryostat Measurement Setup

All measurements were performed in a helium storage dewar with samples in vapor. Two JST connectors were surface-mount soldered to a G10 circuit board with individual break-out traces of silver-plated copper for soldering to the cryostat wiring harness as shown in Fig. 2. The circuit board was clamped to a copper cold finger with a thermally anchored temperature diode. Since the JST connectors have a rated contact resistance, for measurement purposes a tape was modified to allow a 4-point measurement of an $MgB_2$ trace directly on the tape, i.e., an adjacent trace was modified such that the two ends were used for voltage measurements. The resistance of a trace was measured as the sample was slowly cooled in He vapor. Cooling rates of the circuit board setup were less than 0.1 K/s.

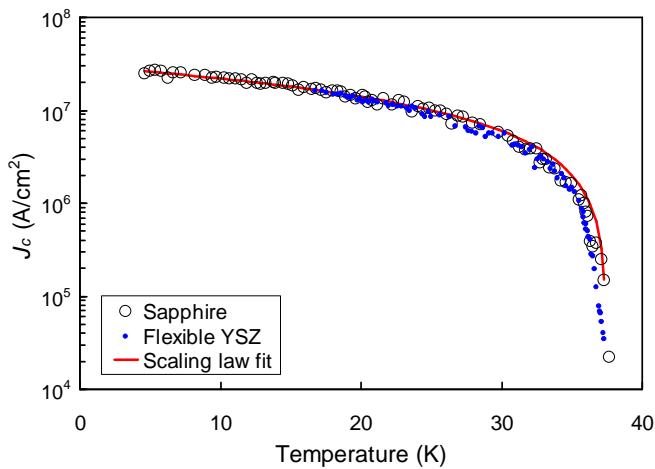

Fig. 4. $J_c(T)$ for 100 nm × 2 µm $MgB_2$ microbridges on $r$-plane $Al_2O_3$ and flexible YSZ. The open circles indicate films on $r$-plane $Al_2O_3$, and the solid dots are films on flexible YSZ. The line is a scaling law fit with $J_o = 30$ $MA/cm^2$.

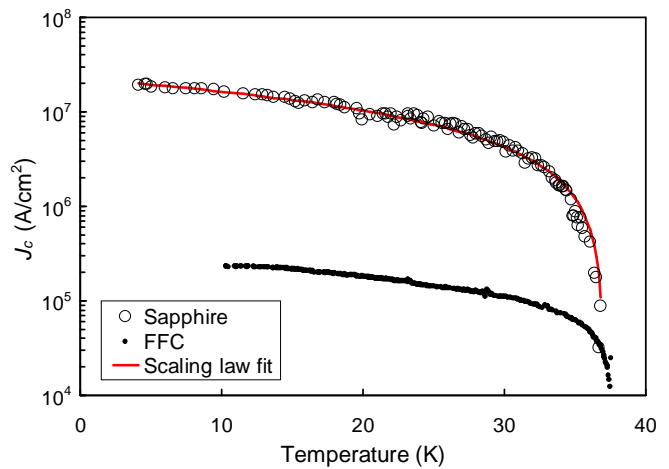

Fig. 5. $J_c(T)$ for 170-nm-thick-$MgB_2$ deposited at 550 ºC. The solid dots are for a 350 µm wide × 10 cm long $MgB_2$ trace on flexible YSZ measured through a JST connector (Fig. 2). The open circles are for a 2-µm-wide-$MgB_2$ microbridge on $r$-plane $Al_2O_3$ (co-deposited with the FFC). The line is a scaling law fit with $J_o = 22.5$ $MA/cm^2$. The temperature shown for the FFC is measured in the center of back of the circuit board, and hence does not accurately represent the temperature of the tape. In addition, the temperature may easily vary along the length of the tape.

### B. $MgB_2$ Thin Films on 1 cm × 1 cm Samples

Fig. 3 shows resistivity as a function of temperature for 100-nm-thick $MgB_2$ films deposited at 550 ºC simultaneously on $r$-plane $Al_2O_3$ and flexible YSZ. It is clear that reactive evaporation of $MgB_2$ on the non-crystalline flexible YSZ substrate produces a high-quality film with $T_c$ and resistivity values comparable to those on $r$-plane $Al_2O_3$. Next we patterned the same $MgB_2$ films into microbridges for $J_c$ measurements. Fig. 4 shows $J_c(T)$ for 2-µm-wide $MgB_2$ microbridges on both substrates. The $J_c(T)$ for $MgB_2$ on flexible YSZ is almost identical to that on $r$-plane $Al_2O_3$. $J_c$ is 5 $MA/cm^2$ at 30 K and greater than $10^7$ $A/cm^2$ at 4.2 K. A scaling law fit of the form $J_c(T) = J_o(1-(T/T_c)^m)$ agrees with the $r$-plane data for $J_o = 30$ $MA/cm^2$, $T_c = 37.5$, $m = 1$.

### C. Critical Current Density of an FFC trace

The critical current density of a complete FFC consisting of $MgB_2$ traces on flexible YSZ mated to JST connectors is shown in Fig. 5. $J_c$ at 30 K is approximately $10^5$ $A/cm^2$ corresponding to an $I_c$ of 60 mA per trace. Assuming that $J_c$ can be maintained for thicker films, these $J_c$ values indicate that a 500-nm film would support 150 mA at 30 K for these 350-µm-wide, 10-cm-long lines. Fig. 4 indicates that $J_c$ values measured for short $MgB_2$ microbridges patterned on 1-$cm^2$ flexible YSZ and $r$-plane $Al_2O_3$ are nearly identical. However, the $J_c$ values of the much wider and longer $MgB_2$ traces on flexible YSZ are more than an order of magnitude lower than what we might expect. One caveat of the $J_c(T)$ data shown in Fig. 5 is that the temperature along the length of the $MgB_2$ tape is unknown. The temperature of the measurement for the FFC shown in Fig. 5 is measured in the center of the back of the circuit board upon which the tape is mounted. While the FFC is not thermally anchored to the circuit board (see Fig. 2), we can assume that the temperatures of both are similar due to He vapor cooling. In addition, since the tape is oriented vertically in the He dewar, the temperature along the length is unknown.

Measurements of the $J_c$ values of two different $MgB_2$ traces on the tape were identical. We also cycled the tape between 4.2 K and room temperature, and upon re-cooling we observed no degradation in $J_c$. In addition, between measurements we flexed the tape to a sagitta of length 25.4 mm (6 cm bend radius) in both directions and found that $J_c$ maintained its previous values.

## IV. THERMAL CONDUCTIVITY

To evaluate the utility of these ribbon cables as low-temperature interconnects, we estimate the overall thermal conductance at 4.2 K. $MgB_2$ has been shown experimentally to be a poor thermal conductor below $T_c$ in both single-crystal (10 W/(m·K) at 4.2 K) [14] and polycrystalline samples (0.1-1 W/(m·K)) at 4.2 K) [15], [16]. By contrast, niobium ($T_c = 9$ K) has been shown to have a thermal conductivity of 10-100 W/(m·K) at 4.2 K [17]. For $MgB_2$, we assume that the contribution to the electronic thermal conductivity (dominant component) due to quasiparticle population exponentially decreases with temperature below $T_c$. For 15 electrical traces of size 1 µm × 350 µm × 10 cm, the estimated thermal conductance at 4.2 K is $5 \times 10^{-8}$ W/K for $MgB_2$ and $5 \times 10^{-7}$ W/K for Nb, using the highest and lowest values for the thermal conductivities respectively.

Regarding the thermal properties of flexible substrates, the thermal conductivity of Kapton HN is 0.01 W/(m·K) at 4.2 K [18]. The thermal conductivity of flexible YSZ at room temperature is 1.5 W/(m·K). While we do not possess low-temperature data for the thermal conductivity of flexible YSZ, it most likely exhibits the Debye model of the lattice thermal conductivity dependence at low temperatures ($\kappa \sim T^3$) and hence may be significantly smaller at 4.2 K. Using a substrate size of 1 cm × 0.01 cm × 10 cm, the estimated thermal conductance at 4.2 K is $10^{-7}$ W/K for Kapton and $10^{-5}$ W/K for flexible YSZ. Clearly, the thermal conductance of an $MgB_2$ FFC will be dominated by the substrate.

## V. ENVIRONMENTAL PASSIVATION

We investigated a number of passivation schemes to protect the $MgB_2$ films from water condensation during thermal cycling. For testing purposes, we submerged the films in deionized water at room temperature ($T$ = 20 ºC) and at $T$ = 90 ºC and visually observed qualitative changes in the films. We evaluated two conformal coatings well known for environmental passivation: Parylene and atomic layer deposition (ALD) of $Al_2O_3$. Parylene is a vapor-phase-deposited polymer commonly used as a moisture barrier for electronics and medical devices. ALD $Al_2O_3$ is currently being investigated as an encapsulant for organic light emitting diodes [19].

In deionized water at 20 ºC, we found that an unprotected $MgB_2$ film will delaminate from an $r$-plane $Al_2O_3$ substrate in approximately 12 to 24 hours. Coatings of 500 nm of Parylene-C or 10 nm of ALD $Al_2O_3$, meanwhile, protect the film equally with no discernable visible degradation for a period of 24 hours.

In $T$ = 90 ºC deionized water, an unprotected $MgB_2$ film dissolves in 10 minutes. A film coated with 500 nm of Parylene-C becomes severely degraded (large $mm^2$ dissolved areas) and discontinuous in 25 minutes. A film with 10 nm of ALD $Al_2O_3$ coating becomes discolored but remains continuous after 25 minutes. An additional 45 minutes is required before the film is severely degraded (complete dissolution of the $MgB_2$ film in areas). Thus, it appears the ALD passivation provides superior protection over Parylene in elevated-temperature degradation tests.

## VI. CONCLUSION

Flexible flat cables of $MgB_2$ on flexible YSZ are an appealing alternative to both commercially available Cu-Ni/Kapton and more exotic materials such as niobium/Kapton. We have demonstrated an FFC consisting of 15 signal traces capable of supporting critical currents of greater than 100 mA each at an operating temperature of 30 K. In addition, extremely high density interconnects are a possibility since all traces can be lithographically defined and wirebonded. We are investigating other possible configurations and applications for $MgB_2$ FFCs including coplanar waveguides and multiple rf striplines on flexible YSZ substrates.

ACKNOWLEDGMENT

The authors thank Thomas Prolier and Tsuyoshi Tajima for the ALD $Al_2O_3$ coatings.